\documentstyle[12pt,graphicx,caption2]{article}

\begin{document}

\title{Absorption in the final state in reactions and decays in the medium}
\author{V.I.Nazaruk\\
Institute for Nuclear Research of RAS, 60th October\\
Anniversary Prospect 7a, 117312 Moscow, Russia.*}

\date{}
\maketitle
\bigskip

\begin{abstract}
The role of strong absorption of particles in intermidiate and final states
has been considered. The range of applicability of phenomenological model
of absorption has been studied. This model is nonuniversal. Its applicability
depends on the type of interaction Hamiltonian and matrix element used. We 
also demonstrate that the violation of the unitarity condition can produce a 
qualitative error in the results. The absorption (decay) in the final state 
does not tend to suppress the total process probability as well as the 
probability of the channel corresponding to absorption. This is true for the 
reactions, decays and $n\bar{n}$ conversion in the medium.  
\end{abstract}

\vspace{5mm}
\vspace{1cm}

*E-mail: nazaruk@inr.ru
\newpage
\setcounter{equation}{0}
\section{Introduction}
In [1] it was shown that in field-theoretical and phenomenological models the 
effect of final state absorption acts in the opposite directions. In the
following this problem is considered in detail. We adduce an additional
arguments and study the reasons for this disagreement. This also makes sense 
if one considers that some problems were solved by means of the above-mentioned
phenomenological models only. Also we study the range of applicability of 
phenomenological models.

Phenomenologically, the absorption is described by an optical potential [2].
For illustration, let us consider a free-space decay $a\rightarrow b\bar{n}$,
for example, $\bar{\Lambda }\rightarrow \bar{n} \pi ^0$. For a decay in
nuclear matter we have
\begin{equation}
a\rightarrow b+\bar{n}\rightarrow b+M
\end{equation}
($M$ are the annihilation mesons) because $\bar{n}$ annihilates in a time
$\tau \sim 10^{-24}$ s.

By way of another example we consider the $n\bar{n}$ transitions [3-5] in
nuclear matter followed by annihilation
\begin{equation}
n\rightarrow \bar{n}\rightarrow M.
\end{equation}

The antineutron annihilation should be described by an Hermitian Hamiltonian
${\cal H}_a$. In the phenomenological models
\begin{equation}
{\cal H}_a\rightarrow  {\cal H}=i{\rm Im}U_{\bar{n}}\bar{\Psi }_{\bar{n}}\Psi_
{\bar{n}},
\end{equation}
where $U_{\bar{n}}$ is the optical potential of $\bar{n}$, ${\cal H}$ is the 
phenomenological absorption (annihilation) Hamiltonian. For brevity, ${\rm Re}
U_{\bar{n}}$ will be omitted, except if otherwise noted.

In practice, the absorption and decay are described by a distorted wave [6],
or dressed propagator (see, for example, refs. [7,8]). To study the model as a
whole, one should write the total interaction Hamiltonian ${\cal H}_I$. In
specific calculations the Hamiltonian ${\cal H}$, as a rule, is not adduced.
However, the corresponding terms in the distorted wave or Green function
originate from ${\cal H}$. Due to this, we consider the problem at the level
of the effective (not fundamental) Hamiltonians.

In the case of process (1), the phenomenological model is given by
\begin{equation}
{\cal H}_I={\cal H}_1+{\cal H}_a\rightarrow {\cal H}_1+{\cal H},
\end{equation}
where ${\cal H}_1$ is the Hamiltonian of the free-space decay $a\rightarrow
b\bar{n}$.

The phenomenological interaction Hamiltonian of process (2) is 
\begin{eqnarray}
{\cal H}_I={\cal H}_{n\bar{n}}+{\cal H},\nonumber\\
{\cal H}_{n\bar{n}}=\epsilon \bar{\Psi }_{\bar{n}}\Psi _n+H.c.,
\end{eqnarray}
$\epsilon =1/\tau $. Here ${\cal H}_{n\bar{n}}$ is the Hamiltonian of $n\bar
{n}$ conversion [5], $\tau $ is the free-space ${n\bar{n}}$ oscillation time.
As we will see later, process (2) is an ideal instrument for the study of the
final state absorption and we focus on this process.

On the one hand, model (3) is very useful because it greatly simplifies
the calculation. On the other hand, the Hamiltonian ${\cal H}$ is non-hermitian
and so model (3) is effective. Its applicability range is restricted.

We consider the decay $a\rightarrow b\bar{n}$ and the $n\bar{n}$ transition in 
the medium and elucidate what processes can be described by the effective
Hamiltonians (4) and (5). More complicated processes are considered as well.
In other words, we study the range of applicability of model (3). We also study
the suppression of the processes mentioned above due to final state absorption.
This is a question of principal because the calculations with hermitian and
non-hermitian Hamiltonians give opposite results.

For our purposes it is sufficient to consider the simplest potential $U_{\bar
{n}}=$const. We perform concrete calculations and show that the unjustified
application of a model can produce a qualitative error in the results. This is
primarily true for the total probability of decays, $n\bar{n}$ transitions and
reactions. Formally, models (4) and (5) can lead to an additional suppression of
the total process probability as well as of the probability of the channel
corresponding to absorption in comparison with the calculations with hermitian
${\cal H}_a$ or, similarly, calculations with hermitian ${\cal H}_a$ can tend
to increase the above-mentioned values.

With the substitution $i{\rm Im}U_{\bar{n}}=-i\Gamma _x /2$, where $\Gamma _x $
is the width of some free-space decay $\bar{n}\rightarrow x$, the effective
Hamiltonians (4) and (5) describe the free-space two-step processes: $a
\rightarrow b+\bar{n}\rightarrow b+x$ and $n\rightarrow \bar{n}\rightarrow x$,
respectively. So, when referring to Hamiltonian ${\cal H}$, we keep in mind the
decay as well.

The paper is organized as follows. In sect. 2 the simple but important
statement related to the final state interaction is proven: the opening of a new 
channel leads to increase the total decay probability. It turns out, that for the
$n\bar{n}$ conversion in the medium model (5) contradicts this statement (sect. 
3). The same is also true for the free-space decay (sect. 4). Section 5 shows that 
the reason for this is the non-unitarity of the $S$-matrix and the structure of the
Green function. In this connection we review the origin of the complex self-energy 
$\Sigma $ in quantum electrodynamics (QED) and optical potential theory and point 
out the principal distinctions with respect to the model under study (sects. 5 and 
6). The value and physical meaning of $i{\rm Im}U$ for various ${\cal H}_I$ are 
analyzed as well. In sect. 7, we qualitatively discuss more complicated 
Hamiltonians and matrix elements. Field-theoretical and phenomenological approaches 
are compared in sect. 8. The results are summarized and discussed in sect. 9.

\section{Absorption in the final state}
To clarify the role of final state absorption, we prove a simple
model-independent statement. We consider the decay $a\rightarrow b\bar{n}$ in
the medium. Let ${\cal H}_s$ and ${\cal H}_a$ be the hermitian Hamiltonians
of the scattering and annihilation of $\bar{n}$, respectively. The total
$a$-particle decay probability $W_t$ is $W_t=W_{\bar{n}}+W_a$, where $W_
{\bar{n}}$ and $W_a$ are the probabilities of finding an antineutron and
annihilation mesons $M$, respectively (see fig. 1).

\begin{figure}[h]
  {\includegraphics[height=.3\textheight]{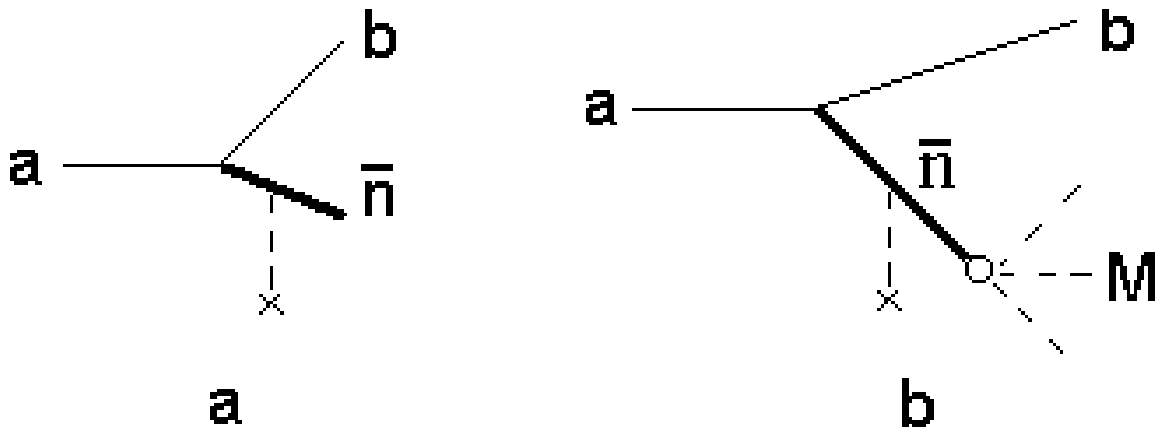}}
  \caption{(a) The decay $a\rightarrow b+\bar{n}$ in the medium without
considering annihilation. (b) The same as (a), but with annihilation.
The antineutron annihilation is illustrated by a circle.}
\end{figure}

Let ${\cal H}_a=0$ and ${\cal H}_s\neq 0$, then $W_t=W_{\bar{n}}$ (see fig. 1a). 
Now, let us turn on the perturbation ${\cal H}_a$ (see fig. 1b). In the lowest
order in ${\cal H}_a$ we have
$$
W_t({\cal H}_s+{\cal H}_a)=W_{\bar{n}}({\cal H}_s+{\cal H}_a)+W_a({\cal H}_s+
{\cal H}_a)>W_{\bar{n}}({\cal H}_s+{\cal H}_a)=W_{\bar{n}}({\cal H}_s+0)=
W_t({\cal H}_s+0).
$$
Thus,
\begin{equation}
W_t({\cal H}_s+{\cal H}_a)>W_t({\cal H}_s+0).
\end{equation}
In the equality $W_{\bar{n}}({\cal H}_s+{\cal H}_a)=W_{\bar{n}}({\cal H}_s+0)$ 
it was taken into account that only terms of zero order in ${\cal H}_a$ give a
contribution to $W_{\bar{n}}$. Inequality (6) can be written in terms of decay 
widths. We use the probabilities for reasons given in sect. 5.

\begin{figure}[h]
  {\includegraphics[height=.25\textheight]{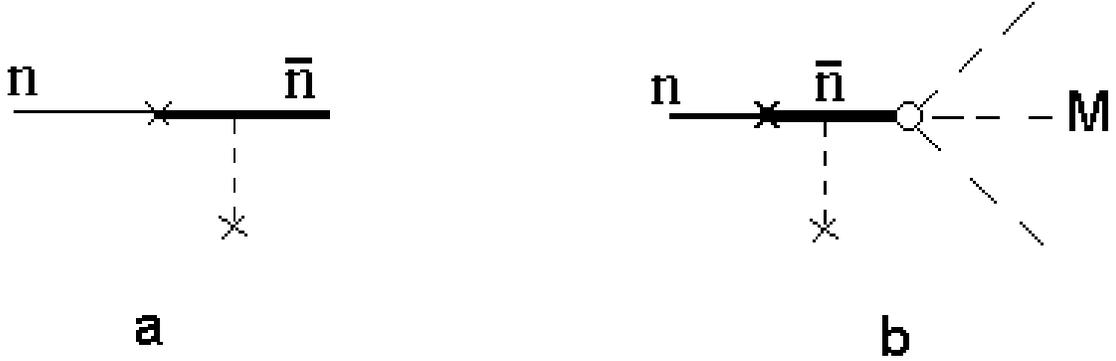}}
  \caption{The same as fig. 1 for the $n\bar{n}$ conversion in the medium.}
\end{figure}

Similar process for the $n\bar{n}$ transition [3-5] in the medium is shown in
fig. 2. Obviously, for this process, inequality (6) is true as well. 

In eq. (6) we can put ${\cal H}_s=0$. Instead of the annihilation we can
consider any process, for example a decay in the final state with the
Hamiltonian ${\cal H}_x$. It is important that there is no interference between 
diagrams $a$ and $b$, and ${\cal H}_x$ is small. The simplest case is given by 
fig. 1, where ${\cal H}_s=0$ and the circle corresponds to the decay. Then in 
the lowest order in ${\cal H}_x$ we have $W_t({\cal H}_1+{\cal H}_x)>W_t({\cal H}
_1+0)$. The expression ${\cal H}_s+{\cal H}_a$ is nothing more than a notation. 
It illustrates the presence of two channels. In principle, the absorption and 
scattering can be described by one and the same interaction Hamiltonian as in QED.

Inequality (6) shows an obvious fact: the opening of a new channel (annihilation) 
leads to increase $W_t$. Obviously, this is generalized to more complicated 
processes: reactions, decays and $ab$ conversion involving final state absorption.

It turns out that models (4) and (5) give the opposite result. This can be
easily shown for $n\bar{n}$ transitions in the medium.

\section{Absorption in the phenomenological model}
In the standard approach (see, for example, refs. [9-13]) the $n\bar{n}$
transitions in the medium are described by Schrodinger equations:
\begin{eqnarray}
(i\partial_t-H_0)n(x)=\epsilon \bar{n}(x),\nonumber\\
(i\partial_t-H_0-V)\bar{n}(x)=\epsilon n(x),\nonumber\\
H_0=-\nabla^2/2m+U_n,\nonumber\\
V=U_{\bar{n}}-U_n={\rm Re}U_{\bar{n}}-i\Gamma /2-U_n,
\end{eqnarray}
$\bar{n}(0,{\bf x})=0$. Here $U_n$ and $U_{\bar{n}}$ are the potentials of $n$ and 
$\bar{n}$, respectively; $\epsilon $ is a small parameter, $\Gamma $ being the 
annihilation width of $\bar{n}$.

For $V=$const. in the lowest order in $\epsilon $ the overall $n\bar{n}$ transition 
probability (the probability of finding an $\bar{n}$ or annihilation
products) in a time $t$ is [13]
\begin{equation}
W_t(t)=1-\mid \!U_{ii}(t)\!\mid ^2=2{\rm Im}T_{ii}(t)-\mid \!T_{ii}(t)\!\mid
^2\approx 2{\rm Im}T_{ii}(t),
\end{equation}
$$
T_{ii}(t)=i(\epsilon /V)^2[1-iVt-\exp (-iVt)],
$$
where $U(t)$ is the evolution operator; $U_{ii}(t)=1+iT_{ii}(t)=<\!n(0)\!\mid
\!n(t)\!>$.

If $V=$const., system (7) has an exact solution. Since $\epsilon $ is 
extremely small, only lowest order in $\epsilon $ is commonly taken into
account. This is a sole approximation made in the calculation of the $T_{ii}$
in the framework of model (5).

At least for small $V$
$$
W_t({\rm Re}V+i{\rm Im}V)<W_t({\rm Re}V+0),
$$
\begin{equation}
{\rm d}W_t/{\rm d}\Gamma <0,
\end{equation}
which contradicts to (6). Indeed, let $\Gamma t\gg 1$. Then
\begin{equation}
W_t(t)=2\epsilon ^2t\frac{\Gamma /2}{({\rm Re}V)^2+(\Gamma /2)^2}\approx
4\epsilon ^2t/\Gamma .
\end{equation}
This is a well-known result [5,11,12] for $n\bar{n}$ transitions in nuclear
matter. If $(\Gamma /2)^2>({\rm Re}V)^2$ (the realistic set of parameters 
fits this requirement), ${\rm d}W_t/{\rm d}\Gamma <0$. At the point ${\rm 
Re}V=0$, ${\rm d}W_t/{\rm d}\Gamma<0$ as well.

In the opposite limiting case $\mid \!Vt\!\mid \ll 1$,
\begin{equation}
W_t(t)=\epsilon ^2t^2(1-\Gamma t/6)
\end{equation}
and we arrive at eqs. (9) again. On the other hand, at small $V$ inequality 
(6) is also valid. Thus (11) contradicts to (6). In model (5) the effect of
absorption acts in the opposite (wrong) direction, which tends to the 
additional suppression of the $n\bar{n}$ transition.

In (6) and (9) physically identical procedures have been done: ${\cal H}_a=
0 \rightarrow {\cal H}_a \neq 0$ and ${\rm Im}V=0 \rightarrow {\rm Im}V \neq
0$, respectively. The results are opposite. Equation (10) shows that the
potential ${\rm Re}V$ suppresses the $n\bar{n}$ transition, which is certainly
correct, however, $\Gamma $ acts in the same direction, which seems wrong.

\begin{figure}[h]
  {\includegraphics[height=.25\textheight]{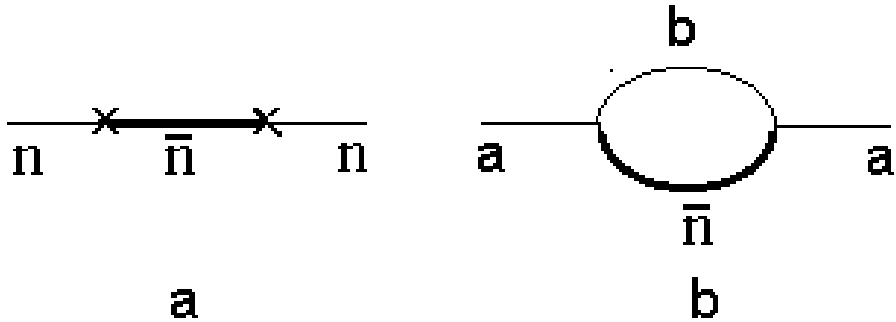}}
  \caption{(a) The on-diagonal matrix element $T_{ii}$ corresponding to the
$n\bar{n}$ transition in the medium. The Hamiltonian of the $\bar{n}$-medium
interaction is given by eq. (3). (b) The same as (a) for the decay 
$a\rightarrow b\bar{n}$.}
\end{figure}

To clarify the structure of (10), we consider the same problem by means of
a diagram technique [1] (see fig. 3a). Here we use the $S$-matrix rather then
an evolution operator. Put $U_n={\rm Re}U_{\bar{n}}=0$ for simplicity.
The Hamiltonian (5) has the form

\begin{equation}
{\cal H}_I=\epsilon \bar{\Psi }_{\bar{n}}\Psi _n+H.c.-i\frac{\Gamma }{2}\bar
{\Psi }_{\bar{n}}\Psi_{\bar{n}}.
\end{equation}
The antineutron propagator $G$ and total process probability $W_t^d(t)$ are
\begin{eqnarray}
G=1/(\epsilon _n-{\bf p}^2/2m+i\Gamma /2)=2/i\Gamma ,\nonumber\\
W_t^d(t)=-2{\rm Im}\epsilon G\epsilon t=\frac {4\epsilon ^2t}{\Gamma },
\end{eqnarray}
where $p=(\epsilon _n,{\bf p})$ is the neutron 4-momentum; $\epsilon _n={\bf
p}^2/2m$. So $W_t^d=W_t$, where $W_t$ is given by (10). The
$\Gamma $-dependence of $W_t$ is conditioned by the propagator $G$. This
fact is common for a 2-tail diagram (fig. 3a) and diagrams with a momentum
transferred $q\neq 0$ (figs. 1b, 3b and 4b).

In this section the standard scheme of calculation has been used. It is based 
on the Hamiltonian (5) and equation $W_t=2{\rm Im}T_{ii}$. This is the sole way 
of calculation of $W_t$ in a one-particle model. For brevity, this model will 
be denoted as model (5). We, thus, see that (5) realized by means of the 
equations of motion or diagram technique contradicts to inequality (6).

\section{Free-space process}
To avoid questions connected with the medium corrections, we consider
the imaginary free-space process
\begin{equation}
n\rightarrow \bar{n}\rightarrow \bar{p}e^+\nu,
\end{equation}
in which the neutron decay is excluded. The hermitian Hamiltonian
is ${\cal H}_I=\epsilon \bar{\Psi }_{\bar{n}}\Psi _n+H.c.+{\cal H}_h^
{\beta }$, where ${\cal H}_h^{\beta }$ is the Hamiltonian of the free-space
$\beta ^+$-decay $\bar{n}\rightarrow \bar{p}e^+\nu$. The corresponding
diagram is shown in fig. 4a.

\begin{figure}[h]
  {\includegraphics[height=.2\textheight]{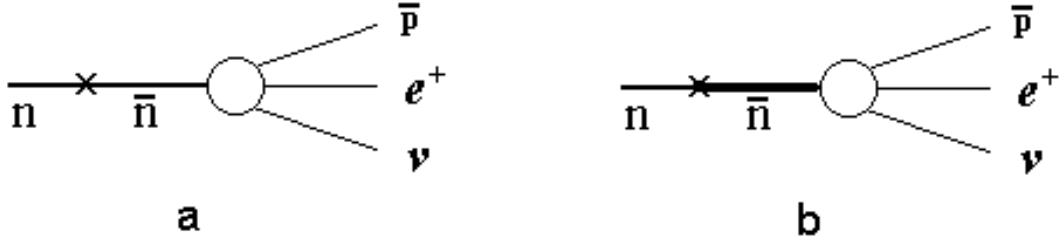}}
  \caption{(a) The free-space process $n\rightarrow \bar{n}\rightarrow \bar{p}e^+
\nu$. (b) The diagram corresponding to the effective amplitude of the process
(14) (see the text).}
\end{figure}

Alternatively, if we want to use model (5), we have
\begin{equation}
{\cal H}_I=\epsilon \bar{\Psi }_{\bar{n}}\Psi _n+H.c.-i\frac{\Gamma _
{\beta }}{2}\bar{\Psi }_{\bar{n}}\Psi_{\bar{n}},
\end{equation}
where $\Gamma _{\beta }$ is the width of the free-space $\beta ^+$-decay
$\bar{n}\rightarrow \bar{p}e^+\nu$. Comparing with (12), it is seen that we
can use all the formulas given above in which $\Gamma =\Gamma _{\beta }$. The
total probability of the free-space $n\bar{n}$ transition $W_t^{\beta }$ is given
by (10) or (13):
\begin{equation}
W_t^{\beta }(t)=-2{\rm Im}\epsilon G\epsilon t\approx 4\epsilon ^2t/\Gamma_
{\beta },
\end{equation}
$\Gamma _{\beta }t\gg 1$. The free-space $n\bar{n}$ conversion is drastically
suppressed by the decay in the final state. Indeed, the free-space $n\bar{n}$
transition probability $W_f$ is $W_f=\epsilon ^2t^2$ (see (11), where $\Gamma
=0$) and correspondingly $W_t^{\beta }(t)/W_f(t)\sim 1/\Gamma _{\beta }t \ll 1$. 
This is clearly wrong because the state the of intermediate $\bar{n}$ (see fig. 
4a) coincides with the final state of the free-space $n\rightarrow \bar{n}$
transition, and so the $\beta ^+$-decay makes no influence on the subprocess of
the $n\bar{n}$ conversion. {\em This is sufficient to reject the model (5)}.

We should make a small comment. The process shown in fig. 4a represents two
consecutive free-space subprocesses.  The speed and probability of the overall
process are defined by those of the slower subprocess. Since $1/\Gamma
_{\beta }\ll t$, the $\beta ^+$-decay can be considered instantaneous: for any
$t_1<t$ the $\beta ^+$-decay probability $W_{\beta }$ is $W_{\beta }(t_1,t)
\approx 1$. Then, the total process probability $W^{\beta}$ is defined by the
speed of the $n\bar{n}$ conversion: $W^{\beta }\approx W_f\sim t^2$ instead of
$W_t^{\beta }\sim t/\Gamma _{\beta }$.

Let us try to compose an effective model which produces $W_t^{\beta }$ through 
the direct calculation of the off-diagonal matrix element. We consider fig. 4b. 
It differs from fig. 4a by the fact that the antineutron is in the potential $V
=-i\Gamma_{\beta }/2$. (Nonsense, of course.) The process amplitude is $M_{eff}
=-\epsilon GM_{\beta}$, $G=-1/V$,  where $M_{\beta}$ is the amplitude of the
$\beta ^+$-decay. For the process width we have $\Gamma_{eff}=\int d\Phi \mid\!
M_{eff}\!\mid ^2/2m=4\epsilon ^2/\Gamma_{\beta }$, which coincides with $W_t^
{\beta }/t$. So we have fixed $<\!f|=<\!\bar{p}e^+\nu|$ and found that the 
effective amplitude of process (14) which produces the result (16), is given by 
fig. 4b.

The fallacy of this model is obvious. Certainly, this is an illustration only,
but structure (16) can be obtained by means of the Green function
$G=-1/V\sim 1/\Gamma $ solely.

In the calculation of the $W_t, W_t^d$ and $W_t^{\beta }$ model (5) is used. The 
result $W_t\sim 1/\Gamma $ is very sensitive to $\Gamma $. However, the $\Gamma 
$-dependence of $W_t$ contradicts to inequality (6). Besides, result (16) is 
unrealistic. Therefore, this model should be revised.

\section{Unitarity and self-energy}
In this and next sections the reasons for disagreement indicated above are studied.
If $\Gamma =0$, system (7) is certainly correct. Consequently, it is necessary to
revise the role of $i{\rm Im}U_{\bar{n}}$. This question has been considered in [1]. 
Taking into account the importance of this problem, we adduce more direct evidence 
using the $U(t)$-operator only. The approach based on the evolution operator is more 
general than the $S$-matrix one, since in this case the time-dependence of the 
process does not need to be $W_t=1-\exp (-\Gamma _tt)$ (see (8)). Also it is 
infrared-free, which is essential for $ab$ transitions [13].

The non-hermiticity of ${\cal H}$ implies that
\begin{equation}
(U(t)U^+(t))_{fi}=\delta _{fi}+\alpha _{fi}(t),
\end{equation}
$\alpha _{fi}\neq 0$, resulting in
\begin{equation}
W_t(t)=\sum_{f\neq i}\mid T_{fi}(t)\mid ^2\approx 2{\rm Im}T_{ii}(t)+
\alpha_{ii}(t)\neq 2{\rm Im}T_{ii}(t)
\end{equation}
because the value of $2{\rm Im}T_{ii}$ is extremely small:
\begin{equation}
2{\rm Im}T_{ii}(t_0)=\frac{4\epsilon ^2t_0}{\Gamma }<10^{-31},
\end{equation}
where the standard set [1, 14-16] of parameters $\epsilon $, $t_0$ and
$\Gamma $ has been used. We thus see that (8) is invalid.

For the $S$-matrix, the conclusions are the same: (a) The basic relation
\begin{equation}
\sum_{f\neq i}\mid T_{fi}\mid ^2\approx 2{\rm Im}T_{ii}
\end{equation}
is inapplicable. (b) The physical meaning of ${\rm Im}\Sigma =-\Gamma /2$ is
uncertain because it is clarified using relation (20). We would like to
emphasize this fact.

On the one hand, the $n\bar{n}$ transition probability is very small (see (19)),
and on the other hand, the term $i{\rm Im}U_{\bar{n}}$ plays a crucial role
because it enters the leading diagram (see (13)). Because of this for the 
problem under study the unitarity of the $S$-matrix is of particular importance.

Thus, the non-hermitian Hamiltonian (3) leads to inverse $\Gamma $-dependence of 
$W_t$ and to the imaginary self-energy. In QED the Green function above threshold 
contains an imaginary self-energy ${\rm Im}\Sigma \neq 0$ as well. However, in the 
case of QED the situation differs principally. ${\rm Im}\Sigma $ is a complicated 
function of parameters of the hermitian Hamiltonian. It appears at higher orders 
in $\alpha $. The width $\Gamma $ makes its appearance after a Dyson summation of 
the relevant self-energies. In order to correctly enforce unitarity, the notation 
of the "running width"
has been introduced.

The importance of unitarity condition is well known [17,18]. Nevertheless,
the non-hermitian models (3)-(5) are frequently used for the reasons given in 
sect. 1. In particular, all existing calculations of $n\bar{n}$ transitions 
in the medium are based on model (5) (see, for example, [13] for future 
references).

With the substitution $i{\rm Im}U_{\bar{n}}=-i\Gamma _x /2$, where $\Gamma _x$
is a width of some free-space decay $\bar{n}\rightarrow x$, the Green function
(13) describes the non-relativistic resonance; Hamiltonians (4) and (5)
correspond to the free-space two-step processes: $a\rightarrow b+\bar{n}
\rightarrow b+x$ and $n \rightarrow \bar{n}\rightarrow x$, respectively. This
is obvious because the absorption can be considered as the decay of a one-particle
state. Formally, in these cases all the results are also true. Nevertheless,
the resonances invite an additional consideration. As far back as 1959, M.
Levy remarked that there does not exist a rigorous theory to which various
phenomenological methods of treating resonances and decays can be considered
as approximations [19]. Attempts have been made at an axiomatic theory [20,21].

The above-mentioned difficulties take place for absorption as well. These 
conceptual problems are beyond the scope of this paper. We deal with concrete models 
(3)-(5) and hence propagator (13) because they are frequently used. As for resonances 
and decays, we only draw the formal analogy between absorption Hamiltonian (3) and phenomenological Hamiltonian of decay $-i\Gamma /2\bar{\Psi }\Psi $.

We also note that decay (14) can be calculated by means of the usual
field-theoretical approach, but the problem should be formulated on the finite
time interval [22] since fig. 4a contains an infrared singularity.

\section{Optical potential}
The problem is not only in the unitarity. It is in the correct description of
the absorption on the whole. In the theory of optical potential $i{\rm Im}U$ is
non-hermitian as well. However, the picture differs principally in this case.
In this section we compare the equation of motion and the problem under study
from the standpoint of the use of an optical potential.

In the case of Schrodinger equation
\begin{equation}
(i\partial_t-H_0-U_{\bar{n}})\bar{n}=0
\end{equation}
the scheme is as follows. Since $U_{\bar{n}}$ is non-hermitian, the condition 
of probability conservation
\begin{equation}
1=\mid U_{ii}\mid ^2+W_{{\rm Sch}}
\end{equation}
is {\em imposed}. Here $W_{{\rm Sch}}$ is the loss of $\bar{n}$ intensity. The
matrix element of evolution operator is found to be
\begin{equation}
U_{ii}(t)=<\bar{n}(0)\mid \bar{n}(t)>=e^{-i{\rm Re}U_{\bar{n}}t}e^{{\rm Im}
U_{\bar{n}}t}.
\end{equation}
From (22) and (23) it is seen that: (a) $W_{{\rm Sch}}\neq 0$ if and only if
${\rm Im}U_{\bar{n}}<0$ (or $\Gamma >0$, when ${\rm Im}U_{\bar{n}}=-\Gamma
/2$). (b) If $\Gamma $ increases, $W_{{\rm Sch}}$ increases as well:
\begin{equation}
\frac{{\rm d}W_{{\rm Sch}}}{{\rm d}\Gamma }=\frac{{\rm d}}{{\rm d}\Gamma }(1-
e^{-\Gamma t})>0.
\end{equation}
This agrees with (6) and is in contradiction with (10), (11) and (16).

The procedure given above is based on two points: 1) In (21) ${\rm Im}U_
{\bar{n}}$ has a clear physical meaning. It is defined by the continuity 
equation corresponding to (21). 2) The additional bound (22) provides the 
probability conservation (unitarization). By means of (21) and (22) $U_
{\bar{n}}$ is fitted to the $\bar{p}$-atom and low energy scattering data.

For more complex problems these requirements, as a rule, are not fulfilled.
We demonstrate this for model (5). The fit of (7) and (8) is impossible since
there are no experimental data. As a result we have (18) with the consequences
considered above. In addition, we try to realize the scheme given for (21).

The coupled eqs. (7) give rise to the following equation
\begin{equation}
(\partial_t^2+i\partial_t(V+2H_0)-H_0^2-H_0V+ \epsilon ^2)n(x)=0.
\end{equation}
According to (8), $n(x)$ is sufficient to get $W_t$.

Even the first step of the scheme described above is not realized: one
cannot get the continuity equation from (25). The $S$-matrix consideration
accomplishes nothing because eq. (20) is inapplicable.

Equations (7), {\it i.e.} model (5), describe only $W_{\bar{n}}$. In this case 
$U_{\bar{n}}$ can be included in the distorted wave of the antineutron 
which is the eigenfunction of eq. (21), and this justifies the model.

\section{Generalization}
If instead of Hamiltonian (5) we take
\begin{equation}
{\cal H}_I={\cal H}_{r,d}+{\cal H},
\end{equation}
where ${\cal H}_r$ and ${\cal H}_d$ correspond to free-space reaction and decay,
respectively, the qualitative conclusions do not change because the heart of
the problem is in the Hamiltonian ${\cal H}$. As an example, let us consider the
decay $a\rightarrow b\bar{n}$ in the medium. Let $\Gamma _{\bar{n}}$ and
$\Gamma _a$ be the widths of decays with $\bar{n}$ and the annihilation mesons in
the final state, respectively;  $\Gamma _t$ is the total decay width,
$\Gamma _t=\Gamma _{\bar{n}}+\Gamma _a$. The corresponding partial decay
probabilities are $W_{\bar{n},a}\approx \Gamma _{\bar{n},a}t$. $W_{\bar{n}}$
and $W_a$ are the same as in sect. 2. To draw the analogy to $n\bar{n}$
transitions, we use the probabilities $W$.

Equations (7) are time-dependent and so the evolution operator has been applied.
For the decays the $S$-matrix is used. In (18) one should replace $T(t)
\rightarrow T$. The interaction Hamiltonian is given by (4). We have
\begin{equation}
\Gamma_t=\frac{2}{T_0}{\rm Im}T_{ii},
\end{equation}
where $T_0$ is the normalization time, $T_0\rightarrow \infty $. The matrix
element $T_{ii}$ is shown in fig. 3b. In principle, the antineutron propagator
in the loop should be calculated through the hermitian Hamiltonian ${\cal H}_a$:
$G=G({\cal H}_a)$. Model (4) means that
\begin{equation}
G({\cal H}_a)\rightarrow G({\cal H})=G(-i\Gamma /2)=-\frac{1}{\hat{p}
_{\bar{n}}-m+i\Gamma/2},
\end{equation}
where $p_{\bar{n}}$ is the antineutron 4-momentum. Obviously, for the matrix
element shown in fig. 3b eq. (18) takes place as well. Relation (20) is
invalid; the physical meaning of ${\rm Im}\Sigma =-\Gamma /2$ is uncertain.

The probability of finding an antineutron $W_{\bar{n}}$ is described by an
off-diagonal matrix element. In the distorted wave impulse approximation the
interaction responsible for the absorption is included in the antineutron
wave function:
\begin{equation}
\bar{n}(x)=\Omega ^{-1/2}e^{-i[({\bf p}^2/2m+i{\rm Im}U)t-{\bf p}{\bf x}]}.
\end{equation}
The corresponding diagram is shown in fig. 1a, where the antineutron state is
described by (29). The wave function $\bar{n}(x)$ is the eigenfunction of
eq. (21), which justifies the using of model (3) in the calculation of 
$\bar{n}(x)$ and $W_{\bar{n}}$.

The probability of finding the annihilation products is obtained from
\begin{equation}
W_a=W_t-W_{\bar{n}}.
\end{equation}
Since eqs. (20) and (27) are inapplicable, $W_t$ and $W_a$ are uncertain.

We thus see that (4) describes only $W_{\bar{n}}$. The result is the
same as for the $n\bar{n}$ transitions considered above. Obviously, in the
strong absorption region $W_{\bar{n}}\ll W_a$ and $W_{\bar{n}}\ll W_t$.

Figures 1b and 2b correspond to absorption in the final state. Model (3) is 
also used for the description of the absorption in the intermediate state (see 
fig. 5). The interaction Hamiltonian of the process shown in fig. 5 has the form
\begin{equation}
{\cal H}_I={\cal H}_1+{\cal H}+{\cal H}_h^{\beta }.
\end{equation}

The quantitative study of models (26) and (31) is subject of a separate
investigation. Here we consider only a qualitative picture. The amplitude
corresponding to fig. 5 is given by
\begin{eqnarray}
T_5=-M_1G(i\Gamma )M_{\beta },\nonumber\\
G(i\Gamma )=[(p_0-q_0-m)-({\bf p}-{\bf q})^2/2m+i\Gamma /2]^{-1}.
\end{eqnarray}
Here $M_1$ and $M_{\beta }$ are the amplitudes of decay (1) and $\beta
^+$-decay, respectively; $p$ and $q$ are the $4$-momenta of particles $a$ and
$b$, respectively; $m$ is the antineutron mass.

\begin{figure}[h]
  {\includegraphics[height=.2\textheight]{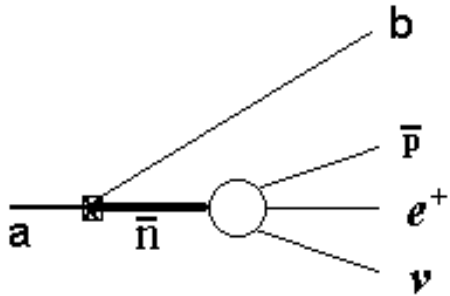}}
  \caption{The decay $a\rightarrow b+\bar{n}\rightarrow b+\bar{p}+e^++\nu$ in the
medium. The bold line signifies the antineutron annihilation in the
intermediate state.}
\end{figure}

The antineutron propagator $G({\cal H}_a)$, calculated through the hermitian
Hamiltonian ${\cal H}_a$, contains the loops. This leads to suppress the
amplitude $T_5$. $G(i\Gamma )$ from (32) acts in the same direction: the
probability of finding the $\beta ^+$-decay products is $W_5\sim \Gamma _
{\beta }/\Gamma ^2$. $W_5$ is less sensitive to model (3) than $W_t$ because
the unitarity condition is not used. In this case at least there is no
qualitative contradiction. (This question needs quantitative consideration.
Most likely model (32) yields too great suppression.)

Below we consider the most realistic case $\Gamma \gg\Gamma _{\beta }$. In
the lowest order in ${\cal H}_h^{\beta }$ the probability of finding an
antineutron $W_{\bar{n}}$ is the same as for Hamiltonian (4). For the
on-diagonal matrix element $T_{ii}$ and total decay probability $W_t$ the
calculation scheme and conclusions are also identical to those for (4)
and (26). The fact that the antineutron propagator in the loop is defined by
${\cal H}_h^{\beta }$ and ${\cal H}$ is not principal because the heart of
the problem is in the $i{\rm Im}U_{\bar{n}}$.

Similarly to (30) we have
\begin{equation}
W_a=W_t-W_{\bar{n}}-W_5.
\end{equation}
Since eq. (20) is inapplicable, $W_t$ and $W_a$ are uncertain.

For model (31) we conclude: (a) $W_{\bar{n}}$ should be described correctly. 
(b) The major decay characteristics $W_t$ and $W_a$ are not described. (c) For 
the process shown in fig. 5 model (32) can be used as a first approximation.

\section{Unitary model}
In sects. 2-4 on the basis of a general reasoning we concluded that in the
phenomenological model the $\Gamma $-dependence of $W_t$ is wrong. Below 
we consider the unitary model and calculate directly the off-diagonal matrix
element by means of the diagram technique.

If $\Gamma t\gg 1$, the probability of finding an annihilation mesons $W_a$ 
is much greater than $W_{\bar{n}}$. However, the phenomenological model describes 
$W_{\bar{n}}$ only. Recall that for the total $n\bar{n}$ transition probability 
the phenomenological model gives $W_t\sim 1/\Gamma $ (see (13)). Since $W_{\bar
{n}}\ll W_t$, $W_a$ depends inversely on $\Gamma $ as well:
\begin{equation}
W_a=W_t-W_{\bar{n}}\approx W_t\sim 1/\Gamma.
\end{equation}
For the processes which are described by Hamiltonians (26) and (31) it is 
sufficient to recall that the $W_t$ and $W_a$ are uncertain for the reasons
given above. In our opinion, with correct consideration of the corresponding
loops we will obtain ${\rm d}W_t/{\rm d}\Gamma <0$, as with (34).

The direct calculation of off-diagonal matrix element gives the inverse
$\Gamma $-dependence ${\rm d}W/{\rm d}\Gamma >0$. Indeed, we consider the
process (1). The $a$-particle and $\bar{n}$ are assumed non-relativistic.
The wave function of the $b$-particle is $\Phi _b(x)=(2q_0\Omega )^{-1/2}\exp 
(-iqx)$, where $q$ is the 4-momentum of the particle. As with ${\cal H}_
{n\bar{n}}$, the decay Hamiltonian is taken in the scalar form ${\cal H}_1=
\epsilon '\bar{\Psi }_{\bar{n}}\Phi ^*_b \Psi _a+H.c.$ and correspondingly
\begin{equation}
{\cal H}_I=\epsilon '\bar{\Psi }_{\bar{n}}\Phi ^*_b \Psi _a+H.c.+{\cal H}_a;
\end{equation}
$\epsilon '$ is dimensionless. 

The process amplitude is given by
\begin{equation}
M_1=-\epsilon '\frac{1}{(p_0-q_0-m)-({\bf p}-{\bf q})^2/2m+i0}M_a.
\end{equation}
Here $M_a$ is the annihilation amplitude, $m$ is the antineutron mass, $p$ is
the 4-momentum of the $a$-particle.

For simplicity assume that $m_b/m\ll 1$, where $m_b$ is the mass of the
$b$-particle. It is easy to estimate the width of decay (1):
\begin{equation}
\Gamma _1\approx \epsilon'^2\Gamma /(2\pi ^2).
\end{equation}
The corresponding decay probability is proportional to $\Gamma $:
\begin{equation}
W_a^h=\Gamma _1t\sim \Gamma .
\end{equation}
The index $h$ signifies that the hermitian Hamiltonian is used.

The width of process (2) is also linear in $\Gamma $ [1]:
\begin{equation}
\Gamma _2\sim \Gamma .
\end{equation}
For the Hamiltonians containing three terms the $\Gamma $-dependence of
$W_a^h$ is the same. Thus,
\begin{equation}
W_a^h\sim W_t^h\sim \Gamma ,
\end{equation}
where $W_t^h$ is the total process probability.

From eqs. (40) and (34), we see that the unitary and non-unitary models lead
to inverse $\Gamma $-dependence of the results. Because of this the
calculations with the hermitian ${\cal H}_a$ can tend to increase $W_a$ and
$W_t$. (See also eq. (12) of ref. [1].)

\section{Summary and conclusion}
We list the consequences of an unjustified use of the model based on eqs. (20)
and (3).

1) Equation (9) contradicts to (6) and (24).

2) Result (16) is unrealistic.

3) $W_t\sim 1/\Gamma $, whereas $W^h_t\sim \Gamma $ (see (34) and (40)).

4) The physical meaning and value of the ${\rm Im}\Sigma $ are uncertain 
(see text below (20)).

Model (3) was adapted to quite definite problems. It is justified for the 
problems described by Schrodinger type equations. It also describes the 
complicated processes (reactions, decays and $n\bar{n}$ transitions) with 
$\bar{n}$ in the final state. (More formally, model (3) can be applied to the 
calculation of $W_{\bar{n}}$ corresponding to the Hamiltonians containing 
several terms (eqs. (26) and (31), for example).) As a first approximation, it 
can be used in the calculation of the diagrams like that shown in fig. 5 with 
$\bar{n}$ in the intermediate state. In these cases $W$ are calculated directly 
without the use of the unitarity condition and the calculation of $T_{ii}$.

In other cases, when the interaction Hamiltonian contains several terms and the
unitarity condition is used (eqs. (8), (13), (27) and (33), for example), model 
(3) is inapplicable. The calculation of the total process probability $W_t$
(and thus $W_a$) corresponding to inclusive reaction, decay or $ab$ transition 
is impossible. The physical meaning of ${\rm Im}\Sigma $ is uncertain. The 
effect of absorption, as a rule, acts in the opposite (wrong) direction, which 
leads to additional suppression. In particular, model (5) gives rise to the 
dramatic suppression of $n\bar{n}$ transitions due to the annihilation in the 
final state, which is wrong. 

This paper also demonstrates the importance of the unitarity condition for any
model of $\Sigma $ [23,24]. The model should by unitary or unitarized.

Finally, we touch upon the result sensitivity to model (3). It is seen from
the Green function (32). The $\Gamma $-dependence is masked by $q$. If
$q\rightarrow 0$ (2-tail) and $m_a=m$, the problem is extremely
sensitive to $\Gamma $: $T_5\sim 1/\Gamma $. Alternatively, in the
phenomenological model the $n\bar{n}$ conversion is described by system (7)
which has an exact solution. For these reasons the $n\bar {n}$ transitions in
the medium are the ideal instrument for the study of the final state absorption.

We also emphasize the following: the absorption (decay) in the final state
(figs. 1b, 2b and 4a, for example) does not lead to suppress the total
process probability as well as the probability of the channel corresponding
to absorption, in contrast to the phenomenological model results. Therefore,
the calculations based on unitary models can tend to increase the
above-mentioned values.\\
\\
The author is grateful to Prof. E. Oset for helpful comments.

\end{document}